\documentclass[12pt]{iopart}

\def\t{\tilde}
\def\ts{\tilde s}
\def\rmd{\mathrm{d}}
\def\rmi{\mathrm{i}}
\usepackage{iopams}
\usepackage{graphicx}

\begin{document}

\title[Propagation-invariant vortex Airy beam whose singular point follows its main lobe]{Propagation-invariant vortex Airy beam whose singular point follows its main lobe}

\author{Masato Suzuki$^1$, Keisaku Yamane$^1$, Takashige Omatsu$^{2,3}$, Ryuji Morita$^1$}

\address{$^1$ Department of Applied Physics, Hokkaido University, Kita-13, Nishi-8, Kita-ku, Sapporo 060-8628, Japan}
\address{$^2$ Graduate School of Advanced Integration Science, Chiba University, 1-33 Yayoi-cho, Inage-ku, Chiba 263-8522, Japan}
\address{$^3$ Molecular Chirality Research Center, Chiba University, 1-33 Yayoi-cho, Inage-ku, Chiba 263-8522, Japan}
\ead{suzuki-masato@eng.hokudai.ac.jp}
\vspace{10pt}
\begin{indented}
\item[]June 2021
\end{indented}

\begin{abstract}
We propose and demonstrate a novel vortex Airy beam which is a superposition of an Airy beam and its laterally sheared beam with a $\pi/2$ phase shift. This new-type of vortex Airy beam exhibits stable propagation dynamics, wherein its singular point closely follows its main lobe, unlike conventional vortex Airy beams. Notably, the orbital angular mode purity of this new vortex Airy beam is up to 10\,\% better than that of a conventional vortex Airy beam. 
We anticipate that this new type of vortex Airy beam, which combines the characteristics of an optical vortex and a \textit{diffraction-free} Airy beam, will facilitate new directions in  applications such as microscopy, material processing and nonlinear optics.
\end{abstract}

\vspace{2pc}

%\submitto{\NJP}%{\JPA}
%\maketitle

\section{Introduction}

An Airy beam is a class of diffraction-free beams which include Bessel and Mathieu beams \cite{Efremidis:19}. The first theoretical investigation of such beams was reported in 1979.  Berry and Balazs showed that a 1D-Airy wave packet is a solution to the potential-free Schr\"{o}dinger equation \cite{doi:10.1119/1.11855}, following this, Besieris \textit{et.\,al.} also suggested a 2D-Airy wave packet as a solution in 1994 \cite{doi:10.1119/1.17510}. After many years, in 2007, the first finite-energy Airy beam was experimentally demonstrated by Siviloglou \textit{et.\,al.} \cite{PhysRevLett.99.213901}. Having unique features like \textit{propagation-invariance} and a \textit{self-accelerating} parabolic trajectory \cite{Efremidis:19,Dholakia2008Nat}, the Airy beam has been utilized in applications including selective plane illumination microscopy (SPIM) \cite{Vettenburg2014,Yang:14,Nylkeaar4817,Corsetti:20}, rapid three-dimensional volumetric imaging \cite{Kozawa2019}, optical coherence tomography \cite{Zhang:19}, material processing \cite{doi:10.1063/1.4745925,Gecevicius:14}, and optical micromanipulation \cite{Baumgartl2008}.

Soon after the experimental demonstration of an Airy beam, a vortex Airy beam was investigated \cite{OVAiryFirst}.  Conventional vortex Airy beams have a characteristic wherein the optical vortex imposed on the main lobe of the beam easily deforms spatially as the Airy beam propagates away from its focal point. This is because the singular point of the main vortex lobe (the main singular point) deviates from the parabolic trajectory of the vortex Airy beam \cite{Dai:10,Dai:11}. It is difficult to make the singular point follow the parabolic course of the Airy beam and as such, this has been seen as a barrier to their use in practical applications. One such promising application is stimulated emission depletion selective plane illumination microscopy (STED-SPIM) \cite{FRIEDRICH2011L43,Scheul:14,Hernandez:20}, which uses the combination of an Airy beam for the excitation beam and a vortex Airy beam for the STED beam. This next-generation of STED microscopy would yield fast and high-resolution imaging with an unparalleled field of view.

In this manuscript, we propose a new vortex Airy beam, whose singular point follows its main lobe. This new vortex Airy beam is composed of two conventional Airy beams which are the laterally- and phase-shifted with respect to one another. Herein, we refer to this as the new-type vortex Airy beam. We present our research as follows; first, we introduce the basic concept of the new-type vortex Airy beam, following which we theoretically examine the propagation dynamics and the orbital angular momentum (OAM) distribution of the beam. We then detail our experimental generation of the new-type vortex Airy beam, and investigate its propagation dynamics. This is followed by discussion and conclusions of our work. 

\section{Theoretical description of vortex Airy beams}
\subsection{Basic concept}

In order to introduce a new-type vortex Airy beam, we start by examining the paraxial equation of diffraction,
\begin{equation}
	\left (2\rmi\partial_{\t z}  + \partial^2_{\t x}  + \partial^2_{\t y} \right) \varphi = 0, 
	\label{eq:npeml}
\end{equation}
where $\ts\!=\!s/s_0\,(s\!=\!x,y,z)$ is a normalized axis, $x_0(=\!y_0)$ is a scaling factor of the transverse plane, $z_0(=\!kx_0^2)$ is a scaling factor of the propagation axis, $k=2\pi n/\lambda$ is the wavenumber with the wavelength $\lambda$ and the refractive index $n$. The electric field envelope of a 2D-Airy beam $\varphi_\mathrm{Airy}$ is a solution to Eq.~(\ref{eq:npeml}) \cite{PhysRevLett.99.213901,Siviloglou:07}:
\begin{eqnarray}
	\varphi_\mathrm{Airy}(\t x, \t y,\t z; \t x_\rmd, \t y_\rmd) = &\prod_{\t s=\t x+\t x_\rmd,\t y + \t y_\rmd} \mathrm{Ai} [\ts -\t z^2/4 +\rmi a_0 \t z]\nonumber\\&\quad\times\exp \left [ a_0 \left ( \ts-\t z^2/2 \right ) -\rmi\frac{\t z}{2}\left ( \frac{\t z^2}{6}-a_0^2 - \ts \right ) \right ],
\end{eqnarray}
where ($\t x_\rmd$,$\t y_\rmd$) gives the lateral constant shift of the Airy beam, $\mathrm{Ai}(\cdot)$ represents the Airy function, and $a_0$ is an exponential truncation factor. 

\begin{figure}[htbp]
\centering\includegraphics[width=10cm,bb=0 0 399 160]{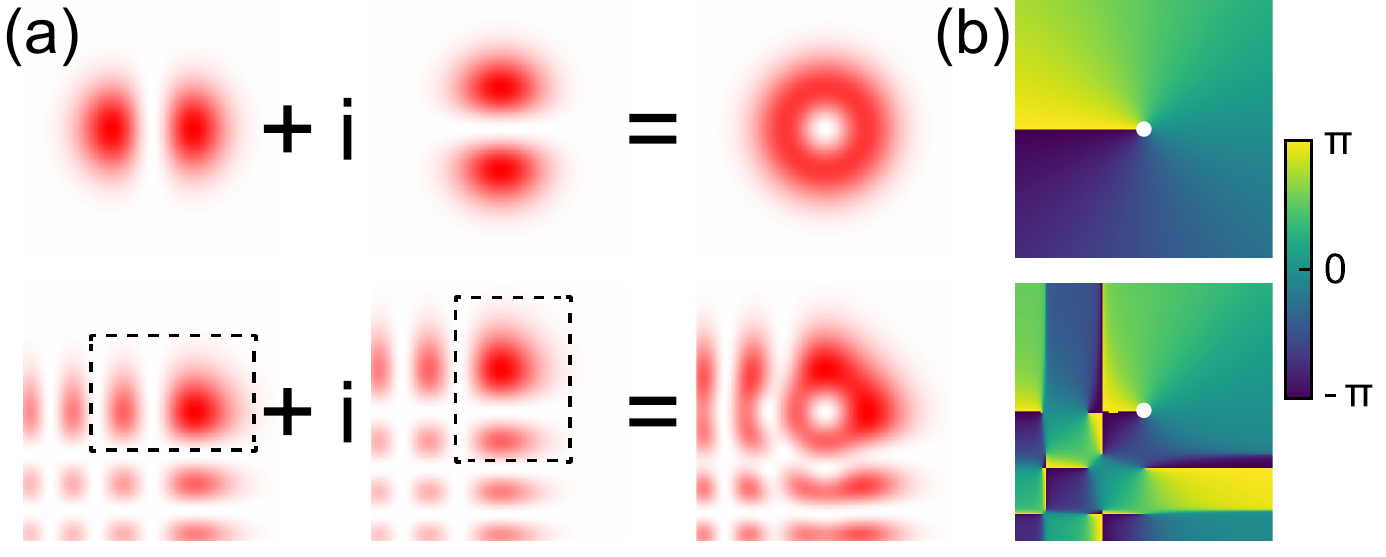}
\caption{(a) The superposition of HG$_{10}$ and HG$_{01}$ modes to form an LG$_{01}$ mode and the superposition of two Airy beams into a new-type vortex Airy beam with a vortex-containing lobe. (b) Plots of the phase distributions of the new-type fields. The white dot indicates the location of the main singular point.\label{fig:sc}}
\end{figure}

The proposed Airy vortex beam is composed of two conventional Airy beams which are superimposed with one another.  As shown in Fig.~\ref{fig:sc}(a), the Laguerre--Gaussian mode with the radial index $p\!=\!0$ and the azimuthal index $\ell\!=\!1$ (LG$_{01}$ mode) can be expressed as the superposition of Hermite--Gaussian modes of order $(i,j)\!=\!(1,0)$ and $(0,1)$ (HG$_{10}$ and HG$_{01}$) with the phase shift of $\pi/2$ \cite{PhysRevA.45.8185}. Now, we can regard parts of an Airy beam as a higher order Hermite--Gaussian mode. For example, the part of an Airy beam which comprises the main lobe and its left neighbor lobe can be approximated as the Hermite--Gaussian mode of order $(i,j)\!=\!(1,0)$. Thus, the superposition of two Airy beams which have a relative phase of $\pi/2$ and a lateral shift, is expected to be a vortex Airy beam carrying $\ell\!=\!+1$ OAM. In fact, there will be an azimuthal phase shift of $2\pi$ around the main singular point (the singular point of the main vortex lobe). In this manuscript, we call this beam a new-type vortex Airy beam with $\ell\!=\!+1$ OAM  (Fig.~\ref{fig:sc}(b)). Similarly, the new-type vortex Airy beam with $\ell\!=\!-1$ OAM can be generated via the superposition of two Airy beams with both a lateral shift and a $-\pi/2$-phase shift.

We here express a new-type vortex Airy beam with $\ell\!=\!\pm 1$ OAM as 
\begin{equation}
	\varphi^\pm_\mathrm{new-type}(\t x, \t y,\t z) = \varphi_\mathrm{Airy}(\t x, \t y,\t z; b_1, b_1') \pm \rmi \varphi_\mathrm{Airy}(\t x, \t y,\t z; b_1', b_1) ,\label{eq:new-typeAiry}
\end{equation}
where $b_k$ and $b'_k$, respectively, represent the $k$th real zeros of $\mathrm{Ai}(\cdot)$ and $\mathrm{Ai}'(\cdot)$ ($b_1\simeq -2.34$ and $b'_1\simeq -1.02$) \cite{NIST:DLMF}.  The total power of the new-type vortex Airy beam in the beam cross section is derived from Parseval's theorem as
\begin{equation}
	\int_{-\infty}^{\infty} \int_{-\infty}^{\infty} |\varphi^\pm_\mathrm{new-type}(\t x, \t y,\t z)|^2 \rmd \t x \rmd \t y = \frac{1}{4\pi a_0}\exp\left ( \frac{4}{3}a_0^3 \right ).
\end{equation}
An $a_0\!=\!0$ new-type vortex Airy beam, while being a non-real solution to the paraxial equation owing to its infinite power, is truly propagation-invariant. We consider this beam to be a perfect vortex Airy beam. As expected, when $a_0\!>\!0$, the new-type vortex Airy beam has a finite energy distribution.

New-type vortex Airy beams with higher-order OAM can be obtained through the superposition of more than two Airy beams. For example, a new-type vortex Airy beam with $\ell\!=\!\pm 2$ OAM is expressed as follows:
\begin{eqnarray}
	\varphi^{\pm 2}_\mathrm{new-type}(\t x, \t y,\t z) = &\varphi_\mathrm{Airy}(\t x, \t y,\t z; b_2', b_1') \pm \rmi \varphi_\mathrm{Airy}(\t x, \t y,\t z; b_1, b_1)\nonumber\\ &- \varphi_\mathrm{Airy}(\t x, \t y,\t z; b_1', b_2').\label{eq:new-typeAiryl=2}
\end{eqnarray}

\subsection{Propagation dynamics of the main singular point}

\begin{figure}[htbp]
\centering\includegraphics[width=\textwidth, bb= 0 0 1406 551]{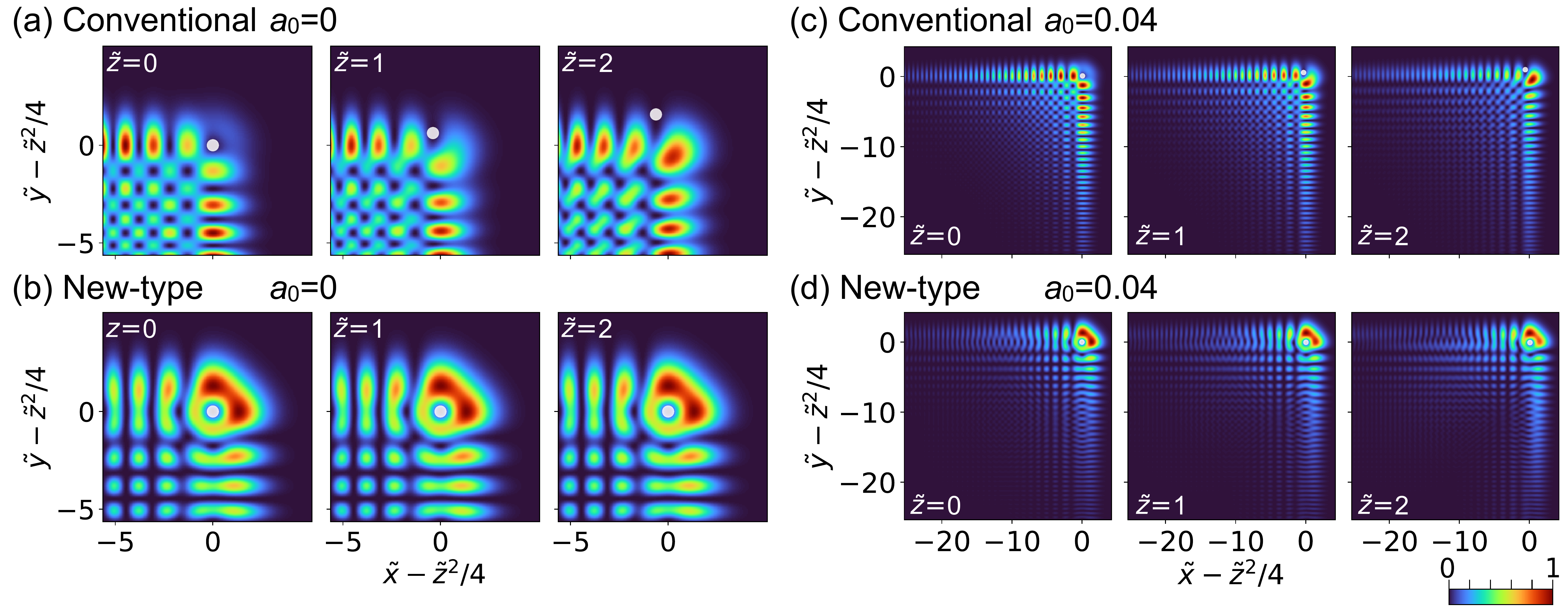}
\caption{Propagation dynamics of new-type vortex Airy beams with $\ell\!=\!+1$ OAM and  conventional vortex Airy beams with $\ell\!=\!+1$ OAM. White dots represent the positions of the $\ell\!=\! +1$ main singular points. (a) Intensity distribution of the perfect conventional vortex Airy beam. (b) Intensity distribution of the perfect new-type vortex Airy beam. (c) Intensity distribution of the $a_0\!=\!0.04$ conventional vortex Airy beam. (d) Intensity distribution of the $a_0\!=\!0.04$ new-type vortex Airy beam. \label{fig:theory}}
\end{figure}

We show that the new-type vortex Airy beam has a vortex lobe which remains stationary (i.e. the position of the singular point does not change) as the distance from the focus ($\t z\!=\!0$) changes. This is in contrast to a conventional vortex Airy beam; the characteristic of which is shown in Fig.~\ref{fig:theory}(a). The deformation of the intensity distribution of the conventional vortex Airy beam is attributed to the main singular point leaving the parabolic trajectory $(\t x,\t y) = (\t z^2/4,\t z^2/4)$ with respect to propagation distance. In the case of the perfect new-type vortex Airy beam (with $a_0\!=\!0$) the main singular point does follow a parabolic trajectory; this can be seen through solution of Eq.~(\ref{eq:new-typeAiry}) as follows:
\begin{equation}
	\varphi^\pm_\mathrm{new-type}(\t x, \t y,\t z;a_0=0) \propto \varphi^\pm_\mathrm{new-type}\left (\t x-\frac{\t z^2}{4}, \t y-\frac{\t z^2}{4},\t z=0;a_0=0 \right ). \label{eq:new-typeAirya0}
\end{equation}
Moreover, the perfect new-type vortex Airy beam maintains the same intensity distribution for any propagation distance; this is shown in Fig.~\ref{fig:theory}(b). 

\begin{figure}[htbp]
\centering\includegraphics[width=11.5cm, bb= 0 0 352 176]{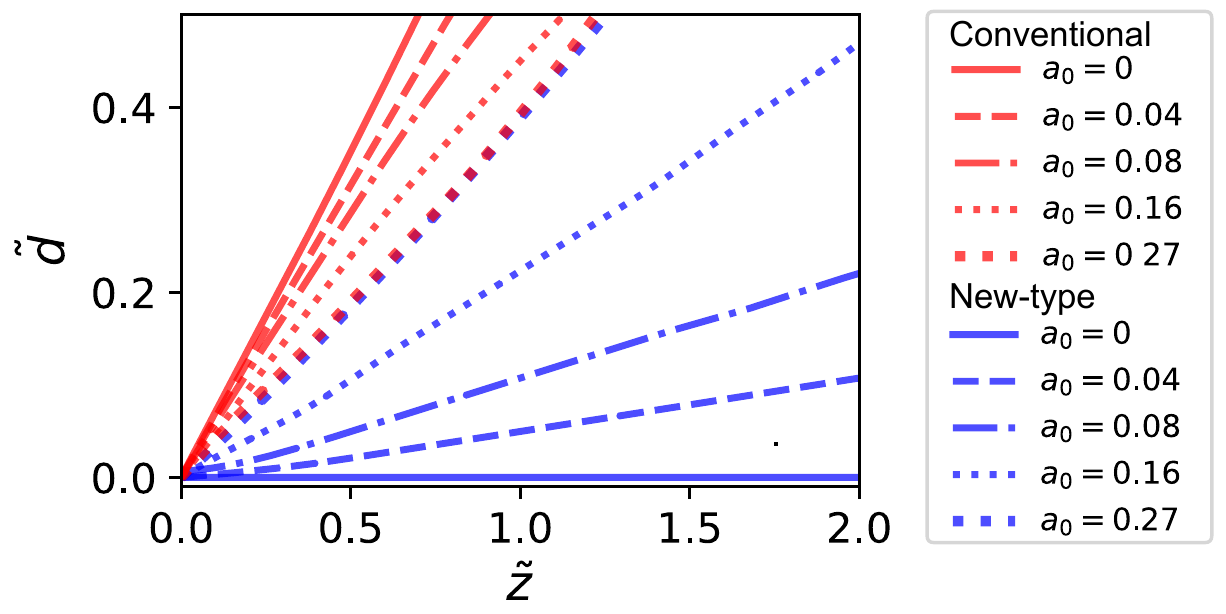}
\caption{The amount of deviation $\t d$ of the main singular point from the parabolic trajectory. \label{fig:theoryd}}
\end{figure}

While the intensity distribution of the perfect new-type vortex Airy beam can be theoretically modeled, it is a non-real solution to the paraxial equation. Real solutions with $a_0\!>\!0$ have also been investigated in this work. In such cases, we observe that there is some deviation of the main singular point from the parabolic trajectory, as the beam propagates away from focus. The amount of deviation $\t d$  (as described in Appendix B)  is however smaller than that observed in a conventional vortex Airy beam in cases where $a_0\!<\!0.27$. Plots of deviation as a function of axial position ($\t z$) for different values of $a_0$ are shown in Fig.~\ref{fig:theoryd}.

When $a_0\!=\!0.27$, the amplitude envelope of the new-type vortex Airy beam decays to $1/e$ times in the main vortex lobe since $b_1 > -a_0^{-1} > b_2(\!\simeq -4.09)$. Usually, the exponential truncation factor is experimentally made small ($a_0 \ll 1$). The smaller $a_0$ is, the smaller the amount of deviation $\t d$ is at the same propagation distance for the new-type vortex Airy beams, in comparison with conventional vortex Airy beams. 
If we consider the case where $a_0\!=\!0.04$, the $1/e$ decay of the new-type vortex Airy beam is in its 26th side lobe since $-a_0^{-1}\simeq b_{27}(\simeq -25.1)$. Here, the  new-type vortex Airy beam preserves the ring shape of the main vortex lobe (as shown in Fig.~\ref{fig:theory}(d)) as it propagates, whereas the main vortex lobe of the conventional vortex Airy beam separates as it propagates (as shown in Fig.~\ref{fig:theory}(c)). 
We find that for $a_0 \t z$ values $\lesssim 0.1$, the ring shape of the main vortex lobe is well-preserved upon propagation and the singular point follows the parabolic trajectory. This is detailed in Appendix B.

\subsection{Orbital angular momentum spectrum}

\begin{figure}[htbp]
%*0.24
\centering\includegraphics[width=12cm, bb= 0 0 461 215]{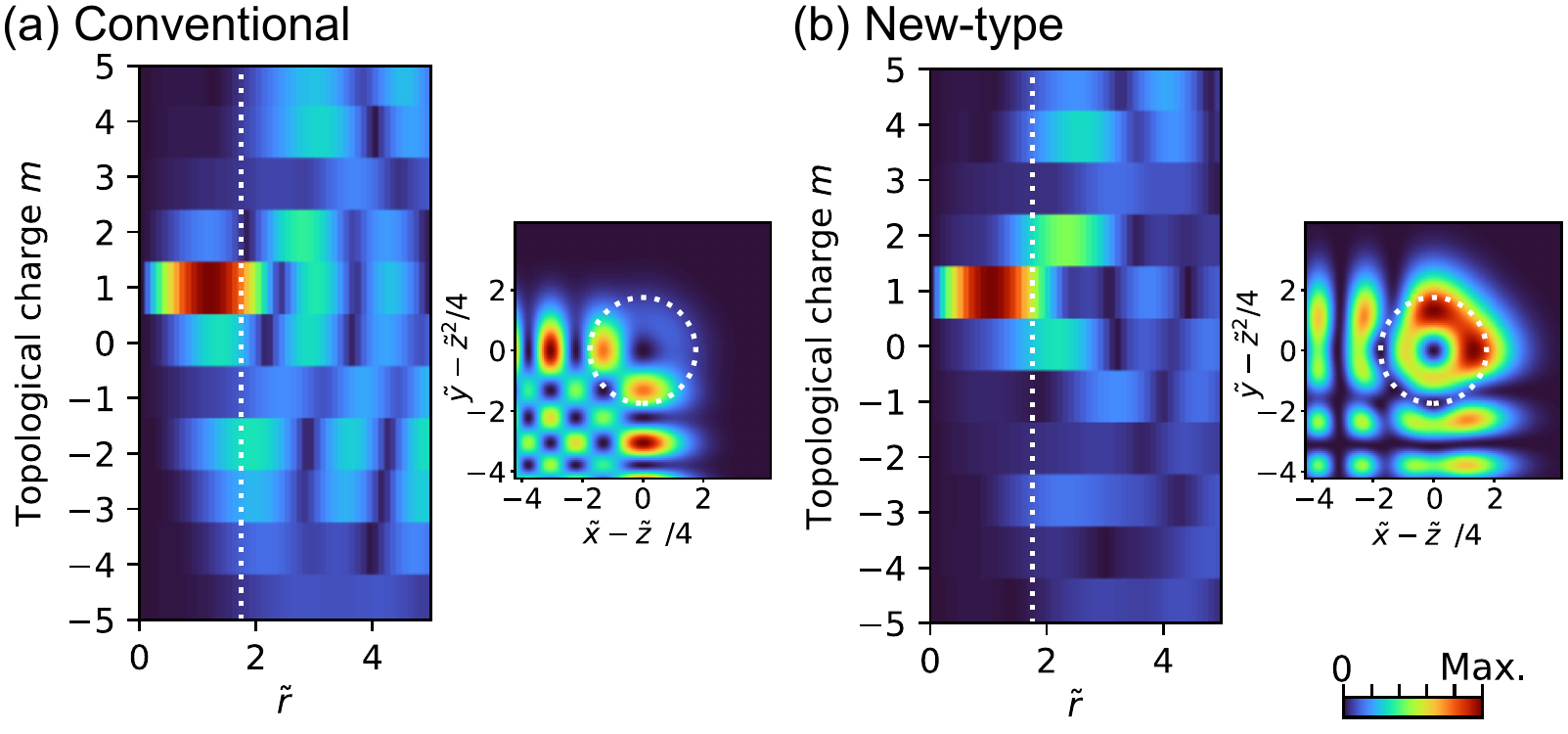}
\caption{The absolute amplitude distributions of OAM spectra of (a) the perfect conventional Airy beam with  $\ell\!=\!+1$ OAM and (b) the perfect new-type vortex Airy vortex beam with $\ell\!=\!+1$ OAM at $\t z\!=\!0$. To clearly resolve the OAM modes, including unwanted modes, the spectral distributions of their absolute amplitude are plotted. The intensity distributions in the real space are also displayed next to the OAM spectra. The white dotted lines indicate $\t r \!=\! b_1-b_2$. \label{fig:tcd}}
\end{figure}

\begin{figure}[htbp]
\centering\includegraphics[width=7cm,bb=0 0 300 208]{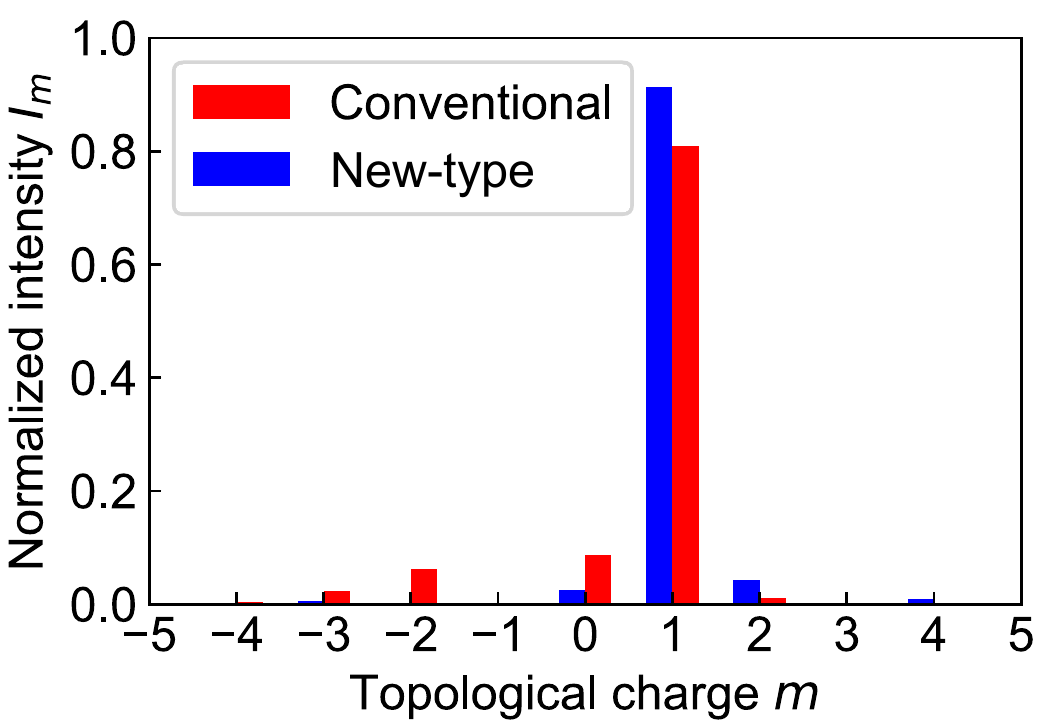}
\caption{OAM spectra of the inner part ($0 \!\le\! \t r \!\le\! b_1-b_2$) of the main vortex lobe of the perfect new-type vortex Airy beam with $\ell\!=\!+1$ OAM and the perfect conventional Airy vortex beam with $\ell\!=\!+1$ OAM at $\t z\!=\!0$. \label{fig:ni}}
\end{figure}

As vortex Airy beams are not symmetric around the main singular point, they carry not only $\ell\!=\!+1$ OAM, but also the other OAM across their profile. Here we examine the OAM spectrum of these vortex Airy beams. The OAM spectrum of the electric field envelope $\varphi$ at $\t z\!=\!0$ is defined by
\begin{equation}
	D_m(\t r) = \frac{1}{2\pi} \int_0^{2\pi} \varphi 
	\left(\t x,\t y, \t z = 0\right ) e^{-\rmi m\phi} \rmd \phi,
\end{equation}
where $m$ represents the topological charge (or OAM in a reduced Planck constant $\hbar$), $\t r \!=\! \sqrt{\t x^2+ \t y^2}$ and $\phi\!=\!\arctan(\tilde y/\tilde x)$ are the normalized radius and the azimuthal angle of the polar coordinates in the transverse plane respectively \cite{Yamane_2014}. 
Figure~\ref{fig:tcd} shows the absolute amplitude distributions of of the OAM spectra of the perfect conventional vortex Airy beam with $\ell\!=\!+1$ OAM and the perfect new-type Airy vortex beam with $\ell\!=\!+1$ OAM. 
In the inner part of the main vortex lobe ($\t r \!\le\! b_1-b_2 \!\simeq\! 1.75$), both of the beams mainly contain $m\!=\!1$ optical vortex modes. Figure~\ref{fig:ni} shows a plot of the OAM spectrum of the inner part of the main vortex lobe of both conventional and new-type vortex Airy beams, where the Intensity ($I_m$) is derived as $I_m\!=\! \int_0^{b_1-b_2}|D_m|^2\t r\rmd \t r / \sum_n \int_0^{b_1-b_2}|D_n|^2\t r\rmd \t r$. The mode purity of the perfect new-type vortex Airy beam is $91\,\%$, while that of the perfect conventional Airy vortex beam is $81\,\%$. Thus, the new-type vortex Airy beam is superior in terms of OAM mode purity as well as beam propagation characteristics. 

\section{Experimental Results and Discussions}

\subsection{Experimental setup}
We experimentally investigated the propagation dynamics of new-type vortex Airy beams, 
generated from an in-house-built Ti:sapphire regenerative amplifier pulsed laser system. 
The output from this laser was horizontally polarized and had a Gaussian spatial profile with a beam radius of 3\,mm. Spectral purity was maintained by passing the beam through a bandpass filter (central wavelength, 800~nm; bandwidth, 5~nm). The laser beam was then incident on a spatial light modulator (SLM) which acted as a phase mask. The details of the phase mask displayed on the SLM can be found in Appendix C. The spatially phase-modulated laser beam was then transformed into a new-type vortex Airy beam by a converging lens ($f\!=\!300$\,mm). Using a CMOS imaging sensor in conjunction with a mechanical stage, we recorded the profiles of the generated vortex Airy beams at different propagation distances, in the vicinity of the focus. For comparison purposes, we generated a conventional vortex Airy beam by implementing a phase mask with a sum of a cubic and a spiral phase distribution on the SLM \cite{Li:18}.

\subsection{Results and Discussion}
\begin{figure}[htbp]
\centering\includegraphics[width=12cm, bb= 0 0 404 295]{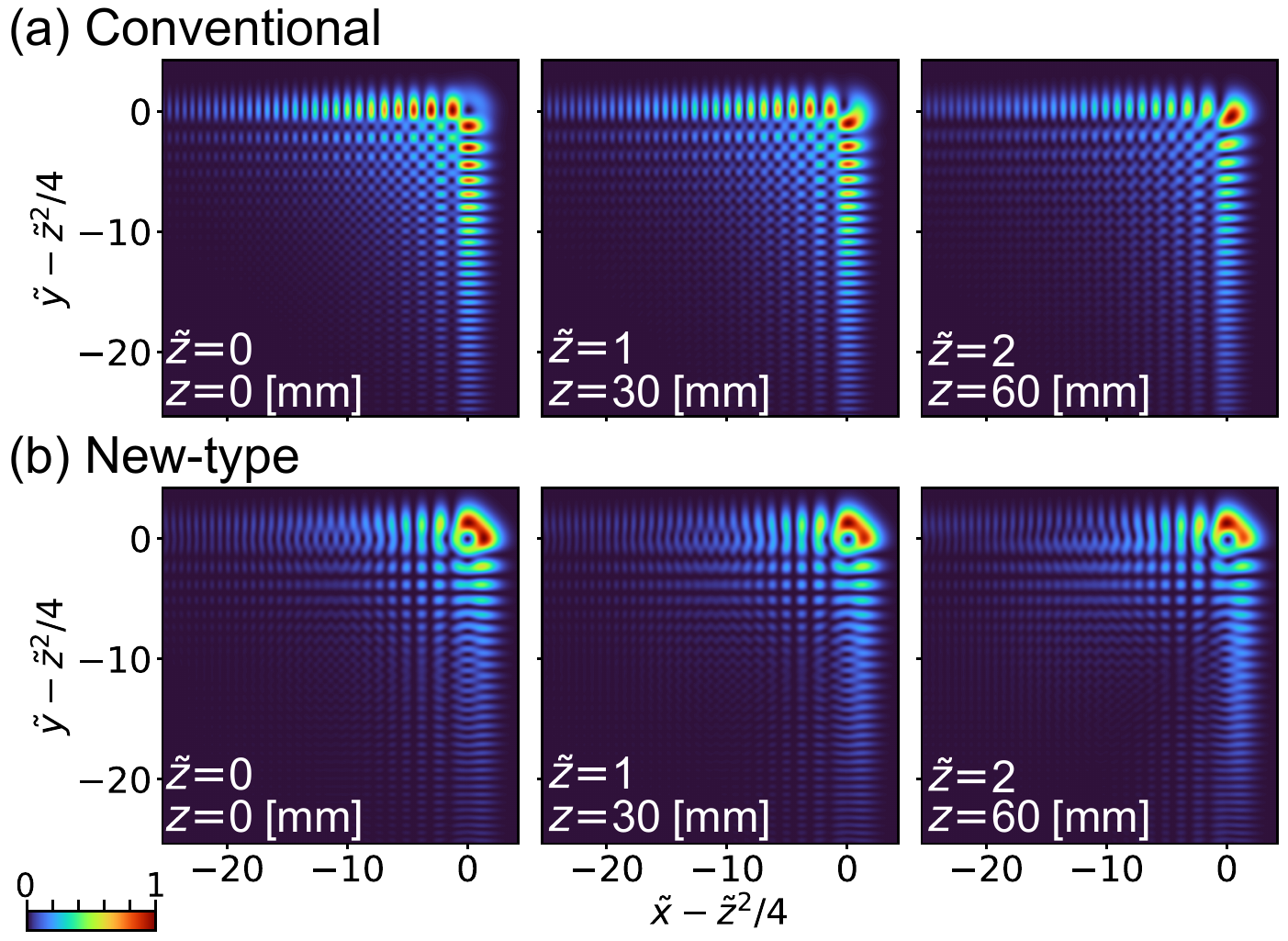}
\caption{Experimentally acquired intensity distributions of (a) the $a_0\!=\!0.04$ conventional vortex Airy beam with $\ell\!=\!+1$ OAM and (b) $a_0\!=\!0.04$ new-type vortex Airy beam with $\ell\!=\!+1$ OAM at $\t z\!=\!0,1,2$.\label{fig:result}}
\end{figure}

Figure~\ref{fig:result} shows the experimental propagation dynamics of a finite-energy new-type vortex Airy beam with  $\ell\!=\!+1$ OAM and a finite-energy conventional vortex Airy beam with  $\ell\!=\!+1$ OAM. The factors of the transverse plane, the propagation axis and the exponential truncation were evaluated to be $x_0\!=\!y_0\!=\!0.06\,$mm, $z_0\!=\!30\,$mm and $a_0\!=\!0.04$, respectively. The propagation dynamics of the conventional vortex Airy beam (Fig.~\ref{fig:result}(a)) and the new-type vortex Airy beam (Fig.~\ref{fig:result}(b)) clearly agree well with the numerical simulations shown in  Figs.~\ref{fig:theory}(c) and \ref{fig:theory}(d), respectively. The main vortex lobe of the finite-energy conventional vortex Airy beam deformed at $\t z=1$ and finally divided into two spots at $\t z=2$, which as mentioned is attributed to the main singular point deviating from the parabolic trajectory. In contrast, the main vortex lobe of the new-type vortex Airy beam maintained its ring shape even at $\t z=2$. This is consistent the main singular point of this new-type vortex Airy beam following the parabolic trajectory (as expected, given $a_0\t z$ was small). 

\begin{figure}[htbp]
\centering\includegraphics[width=12cm,bb = 0 0 563 234]{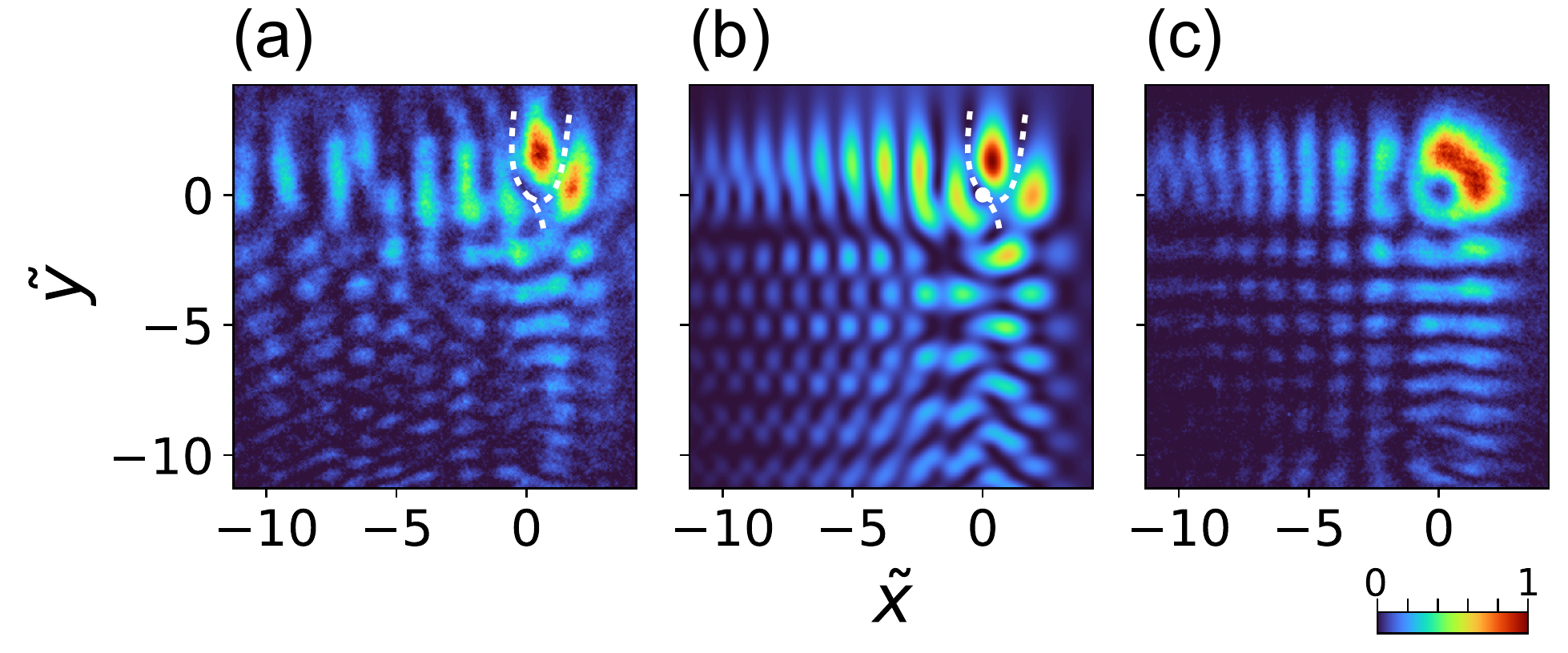}
\caption{(a) Observed image and (b) numerical calculation image of interference pattern at $\t z\!=\!0$. The white dot in (b) represents the main singular point. (c) Observed image without a reference beam at $\t z\!=\!0$. \label{fig:interf}}
\end{figure}

In order to show that a new-type vortex Airy beam has an $\ell\!=\!1$ singular point in the main vortex lobe, we implemented an interference measurement at $\t z=0$ with a reference beam (Fig.~\ref{fig:interf}). This was done using the random mask encoding method \cite{OVAiryFirst} wherein both object and reference beams were simultaneously generated from the same phase mask.  The interference image had a two-pronged fork pattern in the main vortex lobe (Fig.~\ref{fig:interf}(a)) and this was consistent with that predicted via numerical simulation (Fig.~\ref{fig:interf}(b)). Thus, the dominant topological charge of the main vortex lobe was +1 \cite{baranova1981dislocations,Heckenberg:92,Suzuki:18}, which was consistent with the OAM spectrum shown in Fig.~\ref{fig:ni}. These results indicate that these new-type vortex Airy beams comprise a new family of vortex Airy beams.

We comment on future applications of the new-type vortex Airy beams. The dark spot of the singular point is well-preserved in the region of $a_0 |\t z| \lesssim 0.1$, although the shape of the main vortex lobe is not perfectly symmetric especially on propagation.  The new-type vortex Airy beams are expected to possess the \textit{self-healing} properties since they are indeed composed of Airy beams, while it is needed to examine of their \textit{self-healing} properties in future work. Thus, they can be useful for the STED beam in STED-SPIM.  
Light-sheet imaging \cite{Vettenburg2014,Yang:14,Nylkeaar4817,Corsetti:20}, material processing \cite{toyoda2012using,PhysRevLett.110.143603,Nakamura:19} and nonlinear optics \cite{ellenbogen2009nonlinear} may be another fruitful direction.

\section{Conclusion}

In conclusion, we have presented theoretical and experimental investigations into the generation of a new-type vortex Airy beam. These beams exhibit very stable propagation dynamics compared to conventional vortex Airy beams, wherein the position of the singular point within the beam intensity profile does not vary significantly on propagation from focus. This is in contrast to conventional vortex Airy beams which exhibit significant movement of the singular point with beam propagation. We anticipate that the propagation-insensitivity of the singular point in these new-type vortex Airy beams may herald new innovations in applications such as STED microscopy, light-sheet imaging, material processing and nonlinear optics.

%\clearpage

\ack
This work was partially supported by Core Research for Evolutional Science and Technology program (No. JPMJCR1903) of the Japan Science and Technology Agency (JST) and Kakenhi Grants-in-Aid (Nos. JP16H06506, JP17K05069, JP20H02645) from the Japan Society for the Promotion of Science (JSPS).

\appendix

\section{Conventional vortex Airy beam}
In this section, we derive a formula which describes a conventional vortex Airy beam that introduced in Ref.~\cite{OVAiryFirst}. The electric field envelope of the Airy beam
\begin{eqnarray}
	\varphi_\mathrm{Airy}(\t x, \t y,\t z;\t x_\rmd,\t y_\rmd) = &\prod_{\t s=\t x+\t x_\rmd,\t y+\t y_\rmd} \mathrm{Ai} [\ts -\ts_m(z) +\rmi a \t z]\nonumber\\&\quad\times\exp \left [ a \left ( \ts-2\ts_m(z) \right ) -\rmi\frac{\t z}{2}\left ( \frac{\t z^2}{6}-a^2 - \ts \right ) \right ]
\end{eqnarray}
satisfies the normalized paraxial equation of monochromatic electromagnetic waves with  wavenumber $k$
\begin{equation}
	\left (2\rmi\partial_{\t z}  + \partial^2_{\t x}  + \partial^2_{\t y} \right) \varphi = 0, 
	\label{eq:npemlap}
\end{equation}
where $\ts\!=\!s/s_0 (s\!=\!x,y)$ is a normalized axis in the beam cross section, $x_0(\!=\!y_0)$ is a scale factor of the transverse plane, $\t z = z/(kx_0^2)$ is a normalized propagation axis, $\ts_m(z) = \t z^2/4$ defines the lateral shift of the Airy beam, and $a$ is a truncation factor. 

\begin{figure}[htbp]
\centering
\includegraphics[width=.7\linewidth,bb=10 10 442 216]{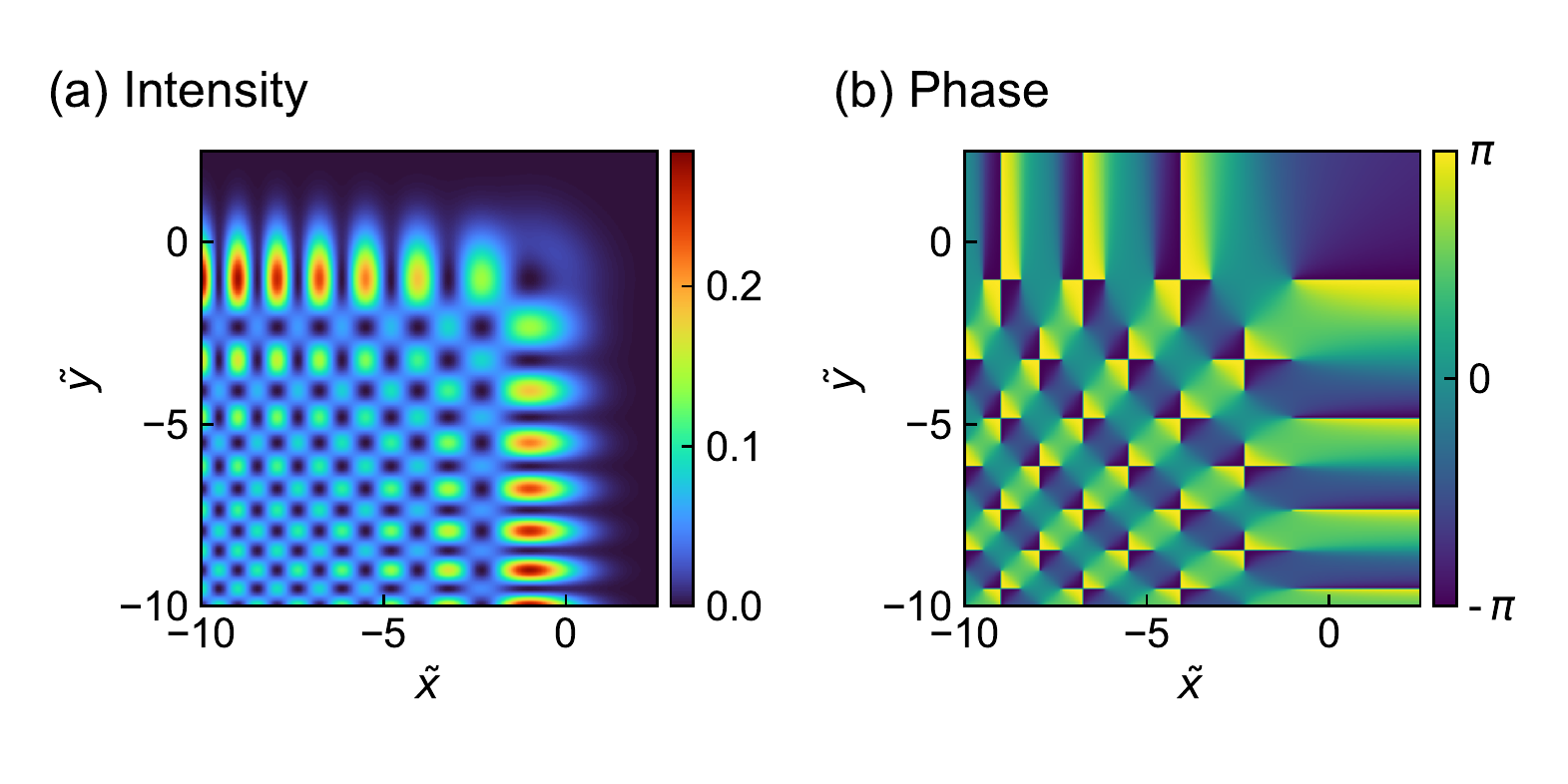}
\caption{(a) Intensity and (b) phase profiles of a vortex array Airy beam $\varphi_\mathrm{array}^+ (\t z\!=\!0; a_0\!=\!0,\t x_\rmd\!=\!0,\t y_\rmd\!=\!0)$.}
\label{fig:vortexarray}
\end{figure}

When an orbital angular momentum operator $\hat L^\pm \equiv \partial_{\t x} \pm i\partial_{\t y}$ \cite{voylar2006} commutes with the operators $\partial_{\t z}$, $\partial^2_{\t x}$, and $\partial^2_{\t y}$ on both sides of Eq.~(\ref{eq:npemlap}), we get 
\begin{equation}
	\left (2\rmi\partial_{\t z}  + \partial^2_{\t x}  + \partial^2_{\t y} \right) \hat L^\pm\varphi = 0, 
	\label{eq:npeml01}
\end{equation}
thus $\hat L^\pm \varphi$ is also a solution of Eq.~(\ref{eq:npemlap}). The explicit form of $\hat L^\pm \varphi_\mathrm{Airy}$ is given by
 \begin{eqnarray}
	\hat L^\pm \varphi_\mathrm{Airy} &= (1\pm \rmi)\left ( a+\rmi\frac{\t z}{2} \right )\varphi_\mathrm{Airy} + \varphi_\mathrm{array}^\pm, \label{eq:convAiryVortex00}\\
	\varphi_\mathrm{array}^\pm &= \left [ \left (\partial_{\t x}\mathrm{Ai} [\t x +\t x_\rmd -\t x_m(z) +\rmi a \t z] \right )\mathrm{Ai} [\t y + \t y_\rmd -\t y_m(z) +\rmi a \t z] \right. \nonumber\\
	&\quad\pm \left. i\mathrm{Ai} [\t x+\t x_\rmd  -\t x_m(z) +\rmi a \t z]\left ( \partial_{\t y}\mathrm{Ai} [\t y + \t y_\rmd -\t y_m(z) +\rmi a \t z] \right )\right ] \nonumber\\
	&\quad\times\prod_{\t s=\t x+\t x_\rmd,\t y+\t y_\rmd}\exp \left [ a \left ( \ts-2\ts_m(z) \right ) -\rmi\frac{\t z}{2}\left ( \frac{\t z^2}{6}-a^2 - \ts \right ) \right ],
	\label{eq:AiryVortexArray00}
\end{eqnarray}
where $\varphi_\mathrm{array}^\pm$ is a vortex array imposed on an Airy beam (Fig.~\ref{fig:vortexarray}). The emergence of a vortex array was reported in the first study of vortex Airy beams \cite{OVAiryFirst}. In general, the orbital angular momentum operator $L^\pm$ adds an orbital angular momentum $\ell = \pm 1$. For example, when the orbital angular momentum operator acts on a Laguerre--Gaussian mode with the radial index $p$ and azimuthal index $\ell$ (LG$_{p\ell}$ mode), the mode will be converted into the LG$_{p(\ell\pm1)}$ mode \cite{voylar2006}. $\hat L^\pm \varphi_\mathrm{Airy}$ (Eq.~(\ref{eq:convAiryVortex00})), however, has the terms of an Airy beam as well as that of a vortex array Airy beam, the characteristic of which causes a degradation of the beam profile with distance from its focus ($\t z\!=\!0$). 
 
Here we show a beam described by Eq.~(\ref{eq:convAiryVortex00}) which is generated by the spatial Fourier transformation of an LG$_{01}$ mode modulated by a cubic phase. The inverse spatial Fourier transform of the Airy beam at $\t z\!=\!0$ is defined as
\begin{eqnarray}
	\mathcal{F}^{-1}[\varphi_\mathrm{Airy}(\t x,\t y,\t z=0;\t x_\rmd,\t y_\rmd)] = \prod_{s=x,y} \exp (-a\t k_s^2)\exp\left [ \frac{\rmi}{3}\{ \t k_s^3-3(a^2 + \t s_\rmd)\t k_s-\rmi a^3\} \right ], \label{eq:invFT00}\nonumber\\
\end{eqnarray}
where $\t k_{s}\!=\! s_0k_{s}$ ($s\!=\!x,y$) is a normalized wavenumber in the transverse plane \cite{OVAiryFirst}. When $a \ll 1$, we get 
\begin{eqnarray}
	\mathcal{F}^{-1}[\varphi_\mathrm{Airy}(\t x,\t y,\t z=0;\t x_\rmd,\t y_\rmd)] = \prod_{s=x,y} \exp (-a\t k_s^2)\exp\left [ \frac{\rmi \t k_s^3}{3}-\rmi \t s_\rmd \t k_s \right ]. \label{eq:invFT01}
\end{eqnarray}
The spatial Fourier transform of Eq.~(\ref{eq:invFT01}) is
\begin{eqnarray}
	\varphi_\mathrm{Airy}(\t x,\t y,\t z=0;\t x_\rmd,\t y_\rmd) \nonumber\\\qquad= \frac{1}{2\pi}\int \!\int\prod_{s=x,y} \exp [-a \t k_s^2]\exp\left [ \frac{\rmi \t k_s^3}{3} \right ] \exp[-\rmi \t k_s(\t s+\t s_\rmd)]\mathrm{d}\t k_x\mathrm{d}\t k_y, \label{eq:FT01}
\end{eqnarray}
which physically means that a Gaussian beam whose beam radius is $a^{-1/2}$ with a cubic phase $(\t k_x^3+\t k_y^3)/3$, propagating through a Fourier lens, generates a finite energy Airy beam at its focus. From Eq.~(\ref{eq:FT01}), a beam described by $\hat L^\pm \varphi_\mathrm{Airy}$ is obtained by Fourier lens transformation of LG$_{01}$ mode beam with a cubic phase,
\begin{eqnarray}
	\hat L^\pm \varphi_\mathrm{Airy}(\t x,\t y,\t z=0;\t x_\rmd,\t y_\rmd) \nonumber\\
	\quad= \frac{1}{2\pi}\int \!\int \prod_{s=x,y} \exp [-a \t k_s^2]\exp\left [ \frac{\rmi \t k_s^3}{3} \right ] \hat L^\pm \exp[-\rmi \t k_s(\t s+\t s_\rmd)]\mathrm{d}\t k_x\mathrm{d}\t k_y\nonumber\\
	\quad=\frac{-\rmi}{2\pi}\int \!\int (\t k_x\pm \rmi \t k_y)\prod_{s=x,y} \exp [-a \t k_s^2]\exp\left [ \frac{\rmi \t k_s^3}{3} \right ] \exp[-\rmi \t k_s(\t x+\t x_\rmd)]\mathrm{d}\t k_x\mathrm{d}\t k_y.\nonumber\\\label{eq:FT02} 
\end{eqnarray}

A conventional vortex Airy beam is usually generated by applying the sum of a cubic phase and a spiral phase to a Gaussian beam \cite{OVAiryFirst,Dai:10,Li:18}. We assume that the radius of the Gaussian beam is $\t w_0$. When $\t w_0 = (2/a)^{1/2}$, $\sim96$\,\% of the phase modulated Gaussian beam is the phase modulated LG$_{01}$ mode beam in terms of energy ratio. Most parts of the conventional vortex Airy beam is described by Eq.~(\ref{eq:convAiryVortex00}). Thereby, obtaining the simple expression of conventional vortex Airy beam, we regard $\varphi_\mathrm{conv.}^\pm(\t x,\t y,\t z)$ as $\hat L^\pm \varphi_\mathrm{Airy}(\t x,\t y,\t z; \t x_\rmd\!=\!b_1',\t y_\rmd\!=\!b_1', a\!=\!2a_0)$.

\section{Propagation dynamics of the main vortex lobe of the new-type vortex Airy beam}
\begin{figure}[htbp]
\centering
\includegraphics[width=.7\linewidth,bb=50 70 536 506]{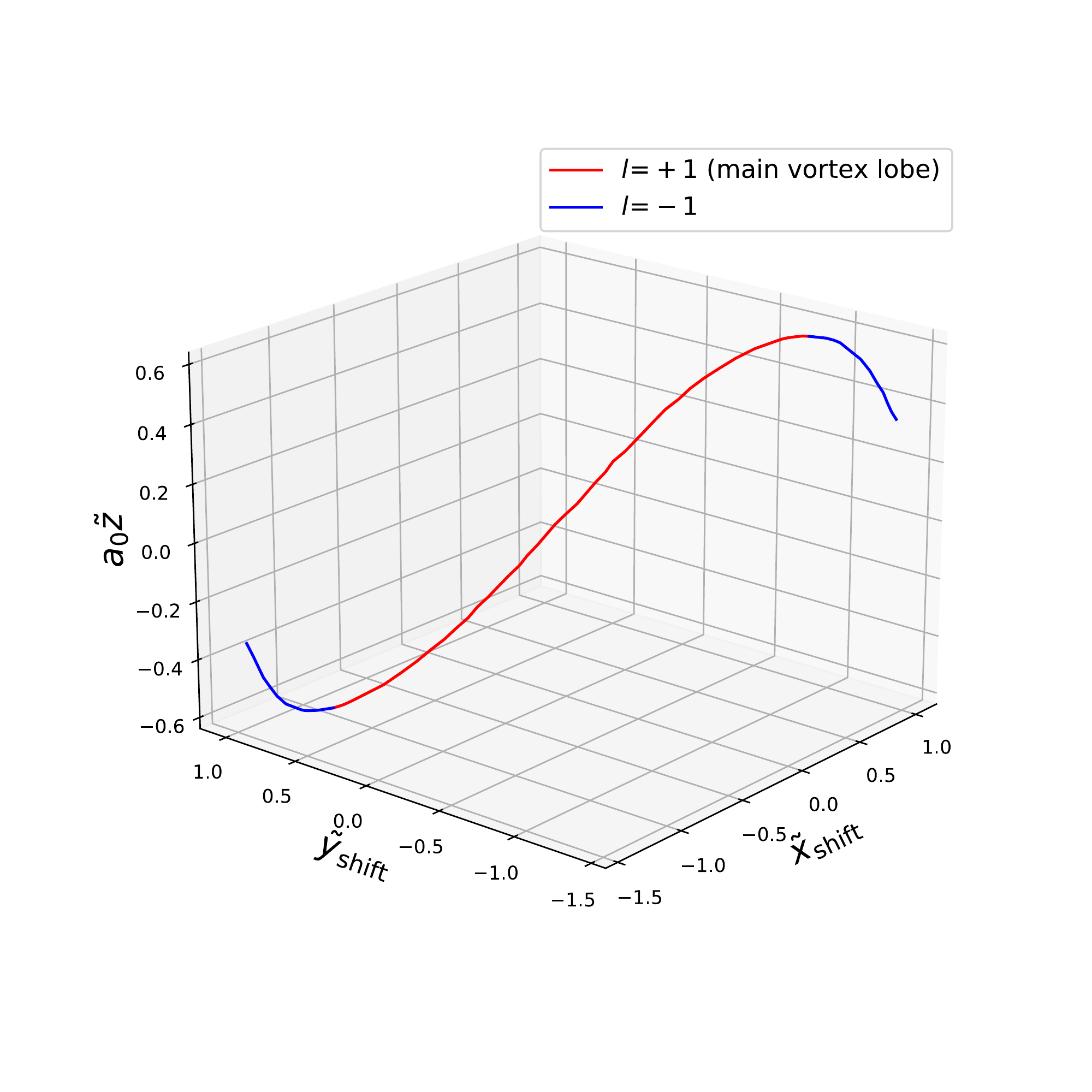}
\caption{Trajectories of singular points of the new-type vortex Airy beam with $\ell\!=\!1$ OAM.}
\label{fig:tsp}
\end{figure}

The position of the singular point $(\t x_\mathrm{sp},\t y_\mathrm{sp})$ of the main vortex lobe of a finite energy new-type beam shifts from the parabolic trajectory $(\t x,\t y)\!=\!(\t z^2/4,\t z^2/4)$. The shift $(\t x_\mathrm{shift},\t y_\mathrm{shift})$ can be numerically fitted by hyperbolic tangent functions.
\begin{eqnarray}
	\left( \begin{array}{c} \t x_\mathrm{shift}(z) \\ \t y_\mathrm{shift}(z) \end{array} \right) &= 
	\left( \begin{array}{c} \t x_\mathrm{sp}(z) - \t x_\mathrm{sp}(0) \\ \t y_\mathrm{sp}(z) - \t y_\mathrm{sp}(0) \end{array} \right) -
	\left( \begin{array}{c} \t z^2/4 \\ \t z^2/4 \end{array} \right) \nonumber\\&= 
	\left( \begin{array}{c} 0.86\cdot \mathrm{arctanh} (1.1\cdot a_0\t z )\\ 0.69\cdot\mathrm{arctanh} (-1.4\cdot a_0\t z ) \end{array} \right)
	\quad(0 \le a_0\tilde z  \lesssim 0.61),\nonumber\\
	\left( \begin{array}{c} \t x_\mathrm{shift}(z) \\ \t y_\mathrm{shift}(z) \end{array} \right) &=
	\left( \begin{array}{c} \t y_\mathrm{shift}(-z) \\ \t x_\mathrm{shift}(-z) \end{array} \right) 
	\quad(-0.61 \lesssim	a_0\tilde z <0).
\end{eqnarray}
$\ell \!= \!\pm 1$ and $\ell \!= \!\mp 1$ singular points appear at $\t z \!\sim\! -0.61/a_0$. The former one is the singular point of the main vortex lobe. This singular point collides with another $\ell \!= \!\mp 1$ singular point, following which these points vanish at $\t z \!\sim\! 0.61/a_0$ (Fig.~\ref{fig:tsp}). 

\begin{figure}[htbp]
\centering
\includegraphics[width=.8\linewidth,bb=0 0 812 553]{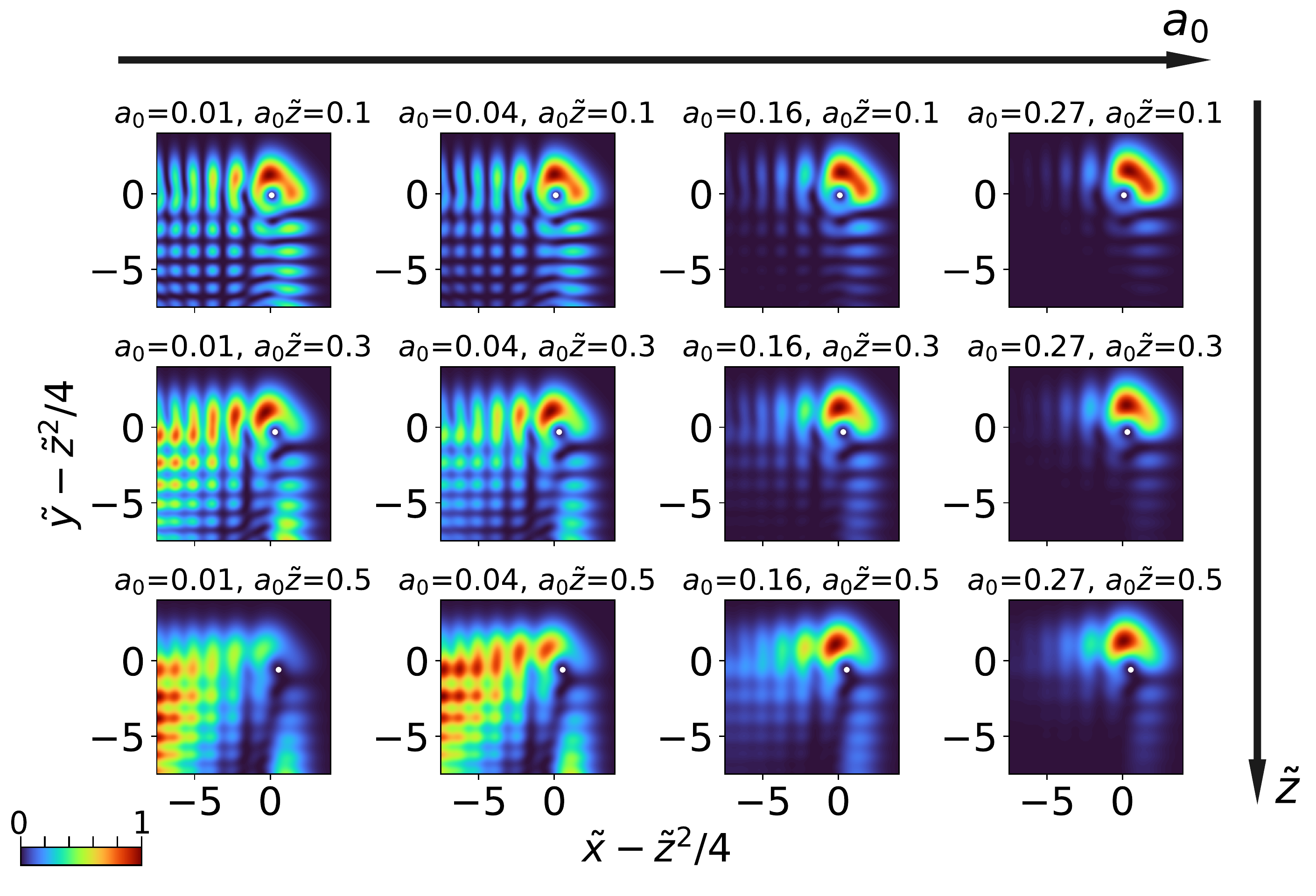}
\caption{Propagation dynamics of $\ell\!=\!1$ $a_0\!=\!0.01, 0.04, 0.16$ and $0.27$ new-type vortex Airy beams at $a_0\t z = 0.1, 0.3$ and $0.5$. White dots represent the position of the main singular point.}
\label{fig:tsp2}
\end{figure}

The main vortex lobe of a finite energy new-type vortex Airy beam deforms with propagation distance. Figure.~\ref{fig:tsp2} depicts propagation dynamics of various new-type vortex Airy beams at $a_0\t z = 0.1, 0.3$ and $0.5$. The shape of the vortex lobe resembles a closed ring at $a_0\t z=0.1$, and it becomes progressively more open as the value of $a_0\t z$ increases, as seen for $a_0\t z=0.3$ and $a_0\t z=0.5$. This characteristics is due to the deviation of the main singular point from the parabolic trajectory ($\t d = \sqrt{\t x_\mathrm{shift}^2+\t y_\mathrm{shift}^2}=0.8$), which is significant with respect to the size of the vortex main lobe.

\section{Phase mask for new-type vortex Airy beam}
\begin{figure}[htbp]
\centering
\includegraphics[width=.7\linewidth,bb=0 0 360 216]{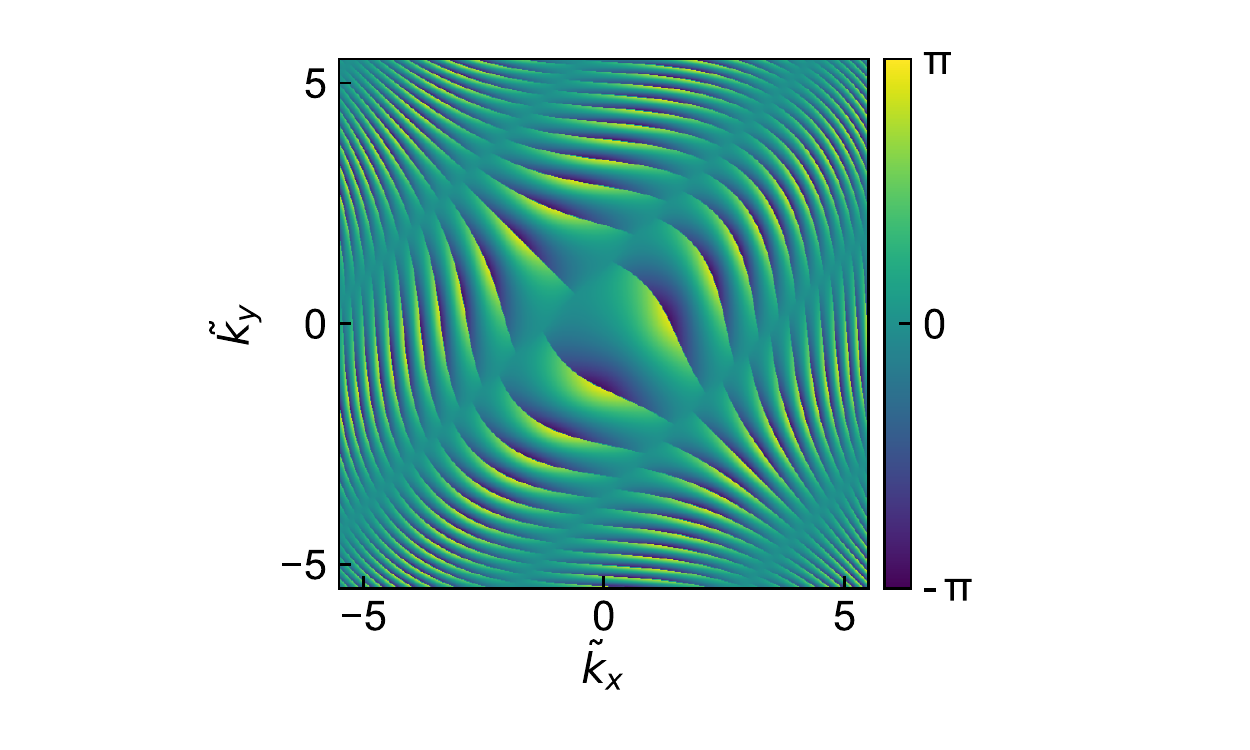}
\caption{Phase distribution of the phase mask without the carrier phase modulation for a new-type vortex Airy beam with $\ell\!=\!1$ OAM.}
\label{fig:phasemask}
\end{figure}

Here we detail the characteristics of the phase mask used to tailor a new-type vortex Airy beam. The Fourier transform of a new-type vortex Airy beam is described by the following expression
\begin{eqnarray}
	\mathcal{F}^{-1}[\varphi_\mathrm{new-type}^\pm(\t x,\t y,\t z=0)] \nonumber\\
	\quad= \mathcal{F}^{-1}[\varphi_\mathrm{Airy}(\t x, \t y,\t z=0; b_1, b_1')] \pm \rmi \mathcal{F}^{-1}[\varphi_\mathrm{Airy}(\t x, \t y,\t z=0; b_1', b_1)]\nonumber\\
	\quad=( \exp [-\rmi (b_1\t k_x+b_1'\t k_y)]\pm \rmi \exp [-\rmi (b_1'\t k_x+b_1\t k_y)])\prod_{s=x,y} \exp (-a_0\t k_s^2)\exp\left [ \frac{\rmi \t k_s^3}{3}\right ]\nonumber\\
	\quad = \sqrt{2}(1\pm\rmi) \sin \left [ \frac{\pi}{4} \mp \frac{(b_1-b_1')(\t k_x-\t k_y)}{2} \right ] \nonumber\\
	\quad \qquad\qquad\qquad\times\prod_{s=x,y} \exp \left [ -a_0\t k_s^2+ \rmi \left ( \frac{\t k_s^3}{3} - \frac{(b_1+b_1')\t k_s}{2}  \right ) \right ],
	\label{eq:FTtAb01}
\end{eqnarray}
where $b_k$ and $b'_k$ represent the $k$th real zeros of $\mathrm{Ai}(\cdot)$ and $\mathrm{Ai}'(\cdot)$ ($b_1\simeq -2.34$ and $b'_1\simeq -1.02$) respectively \cite{NIST:DLMF}. 
We assume the input beam is a Gaussian beam with beam radius $\t w_0\!=\!a_0^{-1/2}$ in the $(\t k_x,\t k_y)$ plane. Since we need both phase modulation and amplitude modulation through a phase mask, we calculated the phase mask pattern by using the Davis's method \cite{Davis:99, Arrizon:07, Ando:09}. Figure~\ref{fig:phasemask} shows the phase distribution that we displayed on the SLM. When $a_0\!=\!0.04$, the radius of the Gaussian beam is $\t w_0\!=\!a_0^{-1/2}\!=\!5$ in the $(\t k_x,\t k_y)$ plane or $w_0\!=\!5/x_0$ in the $(k_x, k_y)\!=\!(\t k_x/x_0,\t k_y/y_0)$ plane. In experiments, the beam radius $w_0$ in the $(k_x, k_y)$ plane is usually a constant value, so $a_0(\!=\!w_0^{-2}x_0^{-2})$ is determined by the scale factor $x_0$.

\section*{References}
\bibliography{bib,sample}

\end{document}